\documentclass[prX,nofootinbib,twocolumn]{revtex4-1}

\usepackage{graphicx}

\usepackage{amsmath,srcltx}
\usepackage{latexsym}
\usepackage{amssymb}    
\usepackage{amsfonts}
\usepackage{multirow}
\usepackage{tikz}
\usepackage[compat=1.1.0]{tikz-feynman}

\newcommand{\be}{\begin{equation}}
\newcommand{\ee}{\end{equation}}
\newcommand{\ba}{\begin{eqnarray}}
\newcommand{\ea}{\end{eqnarray}}

\def\sl#1{\rlap{\hbox{$\mskip 1 mu /$}}#1}

\def\I{\leavevmode\hbox{\small1\kern-3.8pt\norfeynmpmalsize1}}

\begin{document}
\title{Contributions to $Z^0$ decays from a X17 extension of the Standard Model}

\author{D.O.R. Azevedo}
\email{daniel.azevedo@ufv.br}
\affiliation{Universidade Federal de Vi\c cosa (UFV),\\
Departamento de F\'\i sica - Campus Universit\'ario,\\
Avenida Peter Henry Rolfs s/n - 36570-900 -
Vi\c cosa - MG - Brazil.}
\affiliation{Ibitipoca Institute of Physics (IbitiPhys),\\
36140-000 - Concei\c c\~ao do Ibitipoca - MG - Brazil.}
\author{M.L. Bispo}
\email{milena.bispo@ufv.br}
\affiliation{Universidade Federal de Vi\c cosa (UFV),\\
Departamento de F\'\i sica - Campus Universit\'ario,\\
Avenida Peter Henry Rolfs s/n - 36570-900 -
Vi\c cosa - MG - Brazil.}
\affiliation{Ibitipoca Institute of Physics (IbitiPhys),\\
36140-000 - Concei\c c\~ao do Ibitipoca - MG - Brazil.}
\author{O.M. Del Cima}
\email{oswaldo.delcima@ufv.br}
\affiliation{Universidade Federal de Vi\c cosa (UFV),\\
Departamento de F\'\i sica - Campus Universit\'ario,\\
Avenida Peter Henry Rolfs s/n - 36570-900 -
Vi\c cosa - MG - Brazil.}
\affiliation{Ibitipoca Institute of Physics (IbitiPhys),\\
36140-000 - Concei\c c\~ao do Ibitipoca - MG - Brazil.}
\author{J.A. Helay\"el-Neto}
\email{helayel@cbpf.br}
\affiliation{Centro Brasileiro de Pesquisas Físicas (CBPF),\\
Rua Dr. Xavier Sigaud 150, Urca, CEP 22290-180, Rio de Janeiro, Brazil}

\begin{abstract}
The present work proposes a Born-Infeld contribution to the $U(1)_{B-L}$ extension of the Standard Model, where it is introduced associated to the X17 neutral boson. The decay width of the $Z^0$ decay into 3$X$ and 3$\gamma$ processes are computed, based on NA64 and ATLAS experiments data.
\end{abstract}
\maketitle


\textit{Introduction}---. The pursuit of a novel spin 1 particle dates back to the early 80's, pioneered by P. Fayet \cite{fayet1}, where it first appeared as the supersymmetric partner of the goldstino. It was formulated that such a boson could have a mass proportional to an extra very small U(1)-gauge coupling, therefore being present as a very light particle. It has been a topic of research since then. Generally, it is coupled through a linear combination of electromagnetic, $B$ and $L$ (or $B-L$) currents, with the possibility of axial and/or (light) dark matter couplings as well \cite{fayet2,fayet3,fayet4,fayet5,fayet6,fayet7}.

The anomalous decay of the beryllium excited state $^8$Be$^*\to ^8$Be + X \cite{atomki1}, identified by the ATOMKI experiment in 2016, indicated the first possible experimental realization of a fifth gauge interaction appearing in the MeV scale, mediated by a massive X-boson of approximately 17 MeV. It was immediately followed by theoretical propositions to explain the phenomenon \cite{x-model1,x-model2}. Some attempts have been made to explain the anomaly without resorting to an otherwise unknown interaction, primarily within the framework of QCD and nuclear theory, but, as of the present date, all are strongly disfavored, since none of them reached satisfying results \cite{8be-alt,4he-alt}. The results were further corroborated by similar anomalies observed in the nuclear reaction of $^4$He \cite{atomki2} and of $^{12}$C \cite{atomki3}, the last one supporting also the vector character of the new interaction.

Since the ATOMKI initial results, several collaborations have pursued an independent verification of the X17 boson, such as BESIII \cite{besiii}, BelleII \cite{belle-bes,x-boson-search2}, NA64 \cite{x-boson-search1}, BaBar \cite{babar}, with a recent results from the MEG-II \cite{megii} further constraining the branching ratio of the event. Also, the possible interpretation of the X17 as a candidate for a light dark photon, redirects dark matter experiments to search for it, as has been recently done with the PADME experiment \cite{padme1,padme2}, with some upcoming results from its latest run \cite{padme3}, and the New JEDI experiment \cite{newjedi}, which is focused on searching for dark bosons in the MeV range, among others. Contributions coming from other experimental and observational sources \cite{icecube,compobj,anomaly,cern,decays} also allows to further restrict the couplings of the X17 to the Standard Model, as well as new theoretical investigations \cite{libe,hhehh}. For a review of the ATOMKI experiments, the interpretations of their results as new particles and upcoming searches for it, see \cite{review1,review2,review3,review4}.

Considering it as a dark photon candidate, the X-boson could act as a bridge between the dark sector and the visible matter, since it is also coupled to ordinary matter. It can also be kinetically mixed to the photon field or, in the case of a extended electroweak scenario, to the hypercharge sector field. The effective Lagrangian for the X-boson which describes these interactions would be
\begin{equation}
    \mathcal{L} = -\frac{1}{4}F_{\mu\nu}^2 - \frac{1}{4}X_{\mu\nu}^2 + \frac{\chi}{2}F_{\mu\nu}X^{\mu\nu} + \frac{1}{2}m_X^2 X_\mu^2 + J_\mu X^\mu~,
\end{equation}
where $\chi$ is the kinetic mixing parameter. In this sense, the new interaction could be introduced as a $U(1)_{B-L}$ extension of the Standard Model (SM), gauging the global baryonic number minus leptonic number ($B-L$) symmetry, resulting in $SU(3)_C\times SU_L(2)\times U_Y(1)\times U(1)_{B-L}$ as the full gauge group. The new symmetry would have to be spontaneously broken to generate the 17 MeV mass of the X-boson ($m_X$), requiring also the introduction of a Higgs hidden sector in the extended SM \cite{x-boson} or of a Stückelberg mechanism \cite{stueckelberg-mech} to generate the X-boson mass.

The mixing of the X-boson within the abelian sector of the model signals a coupling to the $Z^0$ field, allowing for the search of such particles in processes involving the neutral Z-boson. Therefore, the study of $Z^0$ decays becomes a fruitful field of investigation for this interaction, bearing in mind the proposals for future Z-factories \cite{dm-z-factory} based on $e^- e^+$ colliders like the FCC-ee, and the possibilities it opens for observations of rare decay channels. 

Associated to the further observations of rare $Z^0$ decays is the probing of quartic couplings between neutral gauge bosons, which are ruled out by the Standard Model at tree-level, but could appear already at one-loop order. In our extended SM, this would mean interaction terms appearing between the $Z^0$ and photon but also terms in which the X-boson interacts with the other two. This motivates a non-linear extension of the hypercharge sector, which allows the existence of the anomalous couplings to be treated at tree-level, as well as quartic self-interactions for all neutral gauge bosons, for which there is supporting evidence coming from light-by-light scattering experiments \cite{atlas-gamma-gamma}.

Is this letter, we analyze such a non-linear extension of the SM coupled to the X-boson, which presents neutral boson quartic couplings. This results in dimension 8 effective operators contributing to decay and scattering processes that could indicate novel scenarios of investigation for this new interaction. It is organized as follows: the next section is dedicated to the presentation of the theoretical setup of the model. In the sequence we discuss the decays $Z^0 \to 3 X$ and $Z^0\to 3\gamma$, comparing the latter findings with other results from the literature. Finally, the conclusions is reserved for our final remarks and discussions, as well as the presentation perspectives of future work. We adopt natural units ($\hbar=c=1$) and the Minkowski metric $\eta^{\mu\nu}$= diag$(+---)$ throughout the calculations.

\textit{The model}---. We begin with a discussion of the electroweak sector of the standard model extended by a new abelian $U_{B-L}(1)$ symmetry, such that the gauge group of the theory is $SU_L(2)\times U_Y(1)\times U(1)_{B-L}$, as proposed in \cite{x-boson}. The complete lagrangian can be expressed as
\begin{equation}
\mathcal{L}_{EW-X} = \mathcal{L}_{fermion}+\mathcal{L}_{Higgs}+\mathcal{L}_{gauge}~. 
\end{equation} 
The gauge sector is composed of the fields $A^a_\mu$, $Y_\mu$ and $B_\mu$ associated with $SU_L(2)$, $U_Y(1)$ and $U(1)_{B-L}$, respectively. We introduce a kinetic mixing term $\chi$ into the abelian sector, which results in the following Lagrangian
\begin{eqnarray}
    \mathcal{L}_{gauge} &=&-\frac{1}{4}F^{a}_{\mu\nu}F^{a\mu\nu} -\frac{1}{4}Y_{\mu\nu}Y^{\mu\nu}-\frac{1}{4}B_{\mu\nu}B^{\mu\nu} + \nonumber \\ 
    &+& \frac{\chi}{2}Y_{\mu\nu}B^{\mu\nu}~,
\end{eqnarray} 
with $F^a_{\mu\nu} = \partial_\mu A^a_\nu - \partial_\nu A^a_\mu + g_L\epsilon^{abc}A^b_\mu A^c_\nu $, $B_{\mu\nu} = \partial_\mu B_\nu - \partial_\nu B_\mu$ and $Y_{\mu\nu} = \partial_\mu Y_\nu - \partial_\nu Y_\mu$ being the field-strength of $SU_L(2)$, $U(1)_{B-L}$ and $U_Y(1)$, respectively.

The fermionic sector is defined as
\begin{equation}
    \mathcal{L}_{fermion}=\sum_f i\bar\psi_f\sl{D}\psi_f~,
\end{equation}
where $f$ represents the individual fermion fields and the covariant derivative is 
\begin{equation}
    D_\mu = \partial_\mu -ig_LA^a_\mu T^a-ig_Y Y Y_\mu - ig_{BL} B B_\mu~,
\end{equation}
with $T^a$, $Y$, $B$ and $g_L$, $g_Y$, $g_{BL}$ being the corresponding generators, charges and coupling constants of $SU_L(2)$, $U_Y(1)$ and $U(1)_{B-L}$, respectively. The generators of $SU_L(2)$ satisfy the algebra $[T^a,T^b]=i\epsilon^{abc}T^c$, while the charges $Y$ and $B$ can assume any value.

The particle content, with the respective representation and charges of the fermion fields ($\psi_f$) under each group, is shown on Table \ref{tab:table1}. The index $i$ is a flavor index indicating the fermion families. They are the lepton and quark left chirality $SU(2)$ doublets, $L_i = (\nu^l_i\; l^l_i)^\top$ and $Q_i = (u^l_i \; d^l_i)^\top$, respectively, and the quark and lepton singlets, $u^r_i$, $d^r_i$, $l^r_i$ and $\nu^r_i$, all of them of right chirality. It should be noted the introduction of three right-handed neutrinos $\nu^r_i$, required to guarantee that gauge anomaly is canceled, since the anomaly coefficient is proportional to the sum of the charges ($B_f$) of the fermions under $U(1)_{B-L}$:
\begin{equation}
    \sum_f B_f = 0 ~,
\end{equation}
where the sum runs over all fermion (leptons and quarks) flavors and quark colors.

\begin{table}[h]
\centering
\begin{tabular}{|c|c|c|c|}
\hline
Field & $SU_L(2)$ & $U_Y(1)$ & $U(1)_{B-L}$ \\ 
\hline
$L_i$ & $\underline{\mathbf{2}}$ & $-1/2$  & $-1$ \\
\hline
$Q_i$ & $\underline{\mathbf{2}}$ & $+1/6$  & $+1/3$ \\
\hline
$\nu^r_i$ & $\underline{\mathbf{1}}$ & $0$  & $-1$ \\
\hline
$l^r_i$ & $\underline{\mathbf{1}}$ & $-1$  & $-1$ \\
\hline
$u^r_i$ & $\underline{\mathbf{1}}$ & $+2/3$  & $+1/3$ \\
\hline
$d^r_i$ & $\underline{\mathbf{1}}$ & $-1/3$  & $+1/3$ \\
\hline
$\Phi$ & $\underline{\mathbf{2}}$ & $+1/2$  & $0$ \\
\hline
$\phi$ & $\underline{\mathbf{1}}$ & $0$  & $+2$ \\
\hline
\end{tabular} 
\caption{\label{tab:table1} Particle content of the extended electroweak model and their respective charges under $U_Y(1)$ and $U(1)_{B-L}$.}
\end{table}

Finally, we introduce the Higgs sector, which contains the fields responsible for the electroweak ($\Phi$) and the $U(1)_{B-L}$ ($\phi$) symmetry breaking. The simplest scalar Lagrangian that accomplishes that is
\begin{equation}
    \begin{split}
        \mathcal{L}_{Higgs}&= |D_\mu \Phi|^2 - \lambda \left(\Phi^\dagger \Phi - \frac{\mu^2}{2\lambda}\right)^2 +  |D_\mu\phi|^2 + \\&- \lambda_{BL} \left(\phi^\dagger \phi - \frac{\mu_{BL}^2}{2\lambda_{BL}}\right)^2 - \lambda'|\Phi|^2|\phi|^2~,
    \end{split}
\end{equation}
in which $\mu$, $\mu_{BL}$, $\lambda$, $\lambda_{BL}$ and $\lambda'$ are real parameters, and the covariant derivatives associated to each Higgs fields are given by:
\begin{align}
    &D_\mu \Phi = (\partial_\mu  -ig_LA^a_\mu T^a-ig_Y Y Y_\mu)\Phi ~,\\
    &D_\mu \phi = (\partial_\mu - ig_{BL} B B_\mu)\phi~.
\end{align}

The Higgs scalar fields have non-zero vacuum expectation values (VEV), namely $\langle \Phi \rangle = v = \mu^2 / 2\lambda$ and $\langle \phi \rangle = v_{BL} = \mu_{BL}^2 / 2\lambda_{BL}$, induced by the potentials, responsible for the spontaneous symmetry breaking. The new Higgs $\phi$ must have a VEV scale such that $\langle \phi \rangle \ll \langle \Phi \rangle \simeq 246$ GeV to ensure the light X-boson mass, $m_X\simeq 17$ MeV. This yields the mass matrix for the vector bosons $Y_\mu,A^3_\mu,B_\mu$
\begin{equation}\label{mass-matrix}
    \tilde{M} = \frac{v^2}{4}\left(\begin{array}{ccc}
       g_Y^2  & -g_L g_Y & 0\\
       -g_L g_Y  & g_L^2 & 0\\
        0 & 0 & 16 g_{BL}^2 \frac{v_{BL}^2}{v^2}
    \end{array}\right)~.
\end{equation}

The physical fields of the theory are obtained through a diagonalization procedure, where not only the mass matrix but also the kinetic terms (due to the kinetic mixing term) need to be diagonalized, which results in the following transformation of the neutral gauge fields
\begin{align}
\label{mass-gauge-y}
    &Y_\mu = -\sin{\theta_W}Z_\mu + \cos{\theta_W}A_\mu + \chi\cos^2{\theta_W}X_\mu~,\\
\label{mass-gauge-a}
    &A^3_\mu = \cos{\theta_W}Z_\mu + \sin{\theta_W}A_\mu +\chi\sin{\theta_W}\cos{\theta_W}X_\mu~,\\
\label{mass-gauge-b}
    &B_\mu = -\chi\sin{\theta_W}Z_\mu + X_\mu~,
\end{align}
with the Weinberg angle $\theta_W$ defined by $\sin^2{\theta_W}=0.23$ and $A_\mu$, $Z_\mu$ and $X_\mu$ are the photon, $Z^0$ and the X-boson fields, respectively. Through these transformations, we get the mass eigenstates of the neutral sector, where the photon remains a massless boson, while the $Z^0$ mass receives a contribution from the new gauge field.

The idea proposed here is to modify the hypercharge sector of \cite{x-boson} through a non-linear extension, specifically a Born-Infeld extension, as in \cite{bi-hypercharge1,bi-hypercharge2}:
\begin{equation}
 \mathcal{L}_{BI} = \beta^2 \left(1 - \sqrt{1 - \frac{2}{\beta^2}(\mathcal{F}+\chi\mathcal{H})}\right)~, 
\end{equation}
with $\mathcal{F} = -\frac{1}{4}Y_{\mu\nu}Y^{\mu\nu}$ and $\mathcal{H}=\frac{1}{2}Y_{\mu\nu}B^{\mu\nu}$. The modifications introduced do not compromise the invariant character of the action, since both $\mathcal{F}$ and $\mathcal{H}$ are Lorentz and gauge invariant objects. Since there are no expressions like $\Tilde{Y}_{\mu\nu}Y^{\mu\nu}$, in which the dual tensor $\Tilde{Y}_{\mu\nu}=\frac{1}{2}\epsilon_{\mu\nu\rho\sigma}Y^{\rho\sigma}$ or the similarly defined $\tilde{B}_{\mu\nu}$ field appear, the gauge action is also parity invariant.

In the low energy limit ($\beta\gg|\mathcal{F}|^{1/2},|\mathcal{H}|^{1/2}$), the power expansion of the Born-Infeld Lagrangian reproduces the gauge field terms existent before the non-linear extension, as is shown in equation (\ref{bi-expansion}), and introduces effective operators of dimension 8 and higher which describe interactions that could only appear in loop diagrams before, such as quartic couplings between neutral gauge bosons. 
\begin{equation}
\label{bi-expansion}
    \begin{split}
        \mathcal{L}_{BI} &= \beta^2 \left(1 - \sqrt{1 - \frac{2}{\beta^2}(\mathcal{F}+\chi\mathcal{H})}\right)\\
        &\simeq \mathcal{F} + \chi \mathcal{H} + \frac{1}{2\beta^2}(\mathcal{F}+\chi\mathcal{H})^2 + \mathcal{O}\left(\frac{1}{\beta^4}\right)~.
    \end{split}
\end{equation}
In the limit $\beta\to\infty$, the Born-Infeld Lagrangian reduces to the linear one. Therefore, we shall denote the Lagrangian $\mathcal{L}_{BI}$ as composed of to separate parts, one linear and other non-linear:
\begin{equation}
\label{bi-sum}
    \mathcal{L}_{BI} = \mathcal{L}_{BI-l} + \mathcal{L}_{BI-nl}~.
\end{equation}

If we take into account only the terms of order up to $\mathcal{O}(1/\beta^2)$ in $\mathcal{L}_{BI-nl}$, we get the following list of quartic interactions between neutral bosons:
\begin{equation}
\label{bi-nl}
    \begin{split}
            \mathcal{L}_{BI-nl} &=\mathcal{L}^{4\gamma}+\mathcal{L}^{4Z}+\mathcal{L}^{4X}+\mathcal{L}^{3X-\gamma}+\mathcal{L}^{3X-Z}\\&+\mathcal{L}^{3\gamma-X}+\mathcal{L}^{3\gamma-Z}+\mathcal{L}^{3Z-X}+\mathcal{L}^{3Z-\gamma}+\\&+\mathcal{L}^{2Z-2\gamma}+\mathcal{L}^{2Z-2X}+\mathcal{L}^{2X-2\gamma}\\&+\mathcal{L}^{2\gamma-Z-X}+\mathcal{L}^{2Z-X-\gamma}+\mathcal{L}^{2X-Z-\gamma}~,
    \end{split}
\end{equation}
expressed in terms of the physical fields given by the transformations (\ref{mass-gauge-y})--(\ref{mass-gauge-b}). As stated in the introduction, the expansion above presents all the quartic interactions coming from the SM hypercharge Born-Infeld lagrangian, with a few changes in the constants due to the presence of the new gauge boson, as well as new vertices of the X17 interaction with the $Z^0$ and photon and also its self-interaction term $\mathcal{L}^{4X}$.

An interesting feature is that the photon self-interaction lagrangian $\mathcal{L}^{4\gamma}$ does not receive any contribution from the mixing parameter, since the mass eigenstate remains a massless one, reproducing the expression from the SM (c.f. \cite{bi-hypercharge2} for comparison):
\begin{equation}
    \mathcal{L}^{4\gamma} = \frac{\cos^4{\theta_W}}{32\beta^2}(F^2_{\mu\nu})^2,
\end{equation}
which indicates that any hint for a non-linear theory, in light-by-light scattering, tells nothing about the presence of the extra $U(1)$ sector in this formulation, that is, it cannot distinguish between the non-extended and extended electroweak sectors. It would be an interesting investigation to see if other extended models also present this feature, be it from different non-linear scenarios (such as ModMax \cite{modmax1,modmax2} or Logarithmic \cite{log1,log2,log3}) or from other gauge group extensions and their possible breakings, further motivating the search for non-linear effects in $Z^0$ focused experiments.

In the following, we will discuss the decays $Z^0\to 3\gamma$ and $Z^0\to 3X$, the qualitative differences between them and the contributions arising from the SM extension to the Born-Infeld derived phenomenology of the $Z^0$.

\textit{The $Z^0\to 3X$ decay}---. As we have seen, the expansion of the Born-Infeld Lagrangian generates a series of effective quartic interactions between the photon, the $Z^0$ and the protophobic X-boson. We now proceed with the investigation of those, starting from the decay $Z^0 \to 3 X$, given by the fifth term in (\ref{bi-nl}), which we write explicitly below:
\begin{equation}
\label{L-z3x}
    \mathcal{L}^{3X-Z}= \kappa (X^2_{\mu\nu})(Z_{\rho\sigma}X^{\rho\sigma})~,
\end{equation}
where 
\begin{equation}
\begin{split}
        \kappa&=\frac{\chi^3\sin{2\theta_W}\cos{\theta_W}}{8\beta^2}\times\\&\times\left[\cos^2{\theta_W} - 1 + \chi^2\left(\frac{\cos^4{\theta_W}}{2} - \cos^2{\theta_W}\right)\right]~.
\end{split}
\end{equation}

\begin{figure}[h]
\centering
\begin{tikzpicture}
	\begin{feynman}
		\vertex (a1){\(Z\)};
		\vertex[right=1.5cm of a1] (a2);
		\vertex[right=1.5cm of a2] (a3){\(X\)};
		\vertex[below=1.5cm of a3] (a4){\(X\)};
		\vertex[above=1.5cm of a3] (a5){\(X\)};
		
		\diagram* {
			(a1) -- [boson, edge label=\(p\)] (a2) -- [boson,edge label=\(q_2\) ] (a3),
			(a2) -- [boson, edge label=\(q_3\)] (a4),
			(a2) -- [boson, edge label=\(q_1\)] (a5),
		};
	\end{feynman}
\end{tikzpicture}
\caption{Decay of $Z^0$ into 3$X$.}
\label{graph-z3x}
\end{figure}
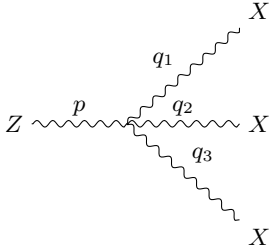

From (\ref{L-z3x}) we can derive the tree-level decay amplitude:
\begin{equation}
\label{decay-amplitude}
    -i\mathcal{M} =  \epsilon_\sigma(p) V^{\mu\nu\rho\sigma}_{Z-3X} \epsilon^*_\mu(q_1) \epsilon^*_\nu(q_2) \epsilon^*_\rho(q_3)~,
\end{equation}
in which $\epsilon_\alpha(k)$ is the polarization vector of the $Z$ and $X$ bosons and the vertex factor $V^{\mu\nu\rho\sigma}_{Z-3X}$, taking into consideration all the possible contractions between the field strengths in (\ref{L-z3x}), reads
\begin{eqnarray}
    V^{\mu\nu\rho\sigma}_{Z-3X} &=& i 4\kappa
        \{ \left[(q_1 \cdot q_2) \eta^{\mu\nu} - q_1^\nu q_2^\mu \right] \left[(q_3 \cdot p) \eta^{\rho\sigma} - q_3^\sigma p^\rho \right] + \nonumber \\
        &+ & \left[(q_1 \cdot q_3) \eta^{\mu\rho} - q_1^\rho q_3^\mu \right] \left[(q_2 \cdot p) \eta^{\nu\sigma} - q_2^\sigma p^\nu \right] + \nonumber \\
        &+ & \left[(q_2 \cdot q_3) \eta^{\nu\rho} - q_2^\rho q_3^\nu \right] \left[(q_1 \cdot p) \eta^{\mu\sigma} - q_1^\sigma p^\mu \right] \}~. \label{vertex_factor}
\end{eqnarray} 
The subscript $Z\rightarrow3X$ of the vertex factor shall be omitted from hereafter, as it should be clear the process from which it derives. We have assumed $p$ flowing into the vertex while each $q_i$ flows out of it, as is expected from a decay process, represented by the Feynman graph in Figure \ref{graph-z3x}.

From (\ref{decay-amplitude}), averaging over the initial state polarization and making use of the completeness relation: 
\begin{equation}
    \sum_r \epsilon_\mu(k,r)\epsilon_\nu(k,r) = -\eta_{\mu\nu} + \frac{k_\mu k_\nu}{m^2}~,
\end{equation}
with $r=-1,0,1$ the polarization states of the massive bosons, we get the unpolarized squared amplitude:
\begin{equation}
\begin{split}
\label{squared-amplitude}
    \langle|\mathcal{M}|^2\rangle &= \frac{1}{3}\sum_i \sum_{r_i}|\mathcal{M}|^2\\
    &=\frac{1}{3}\left(-\eta_{\sigma\delta} + \frac{p_\sigma p_\delta}{m_Z^2}\right)\left(-\eta_{\mu\alpha} + \frac{q_{1\mu} q_{1\alpha}}{m_X^2}\right)\times\\
    &\times\left(-\eta_{\nu\beta} + \frac{q_{2\nu} q_{2\beta}}{m_X^2}\right)\left(-\eta_{\rho\gamma} + \frac{q_{3\rho} q_{3\gamma}}{m_X^2}\right)\times\\
    &\times V^{\mu\nu\rho\sigma}\left(V^{\alpha\beta\gamma\delta}\right)^*~.
\end{split}
\end{equation}

However, the decay amplitude obeys the Ward identity, 
\begin{equation}
    k_{(\mu,\nu,\rho,\sigma)}\mathcal{M}^{\mu\nu\rho\sigma}(k)= 0~,
\end{equation}
for $k_{(\mu,\nu,\rho,\sigma)}$ being one of the momenta ($q_i,p$) of the particles involved in the process and the tensor ${M}^{\mu\nu\rho\sigma}$ defined as the decay amplitude deprived of the polarization vectors $\epsilon_\alpha^{(*)}(k)$.

Therefore, the only non-zero factor from equation (\ref{squared-amplitude}) is
\begin{equation}
    \langle |\mathcal{M}|^2\rangle = \frac{1}{3}V^{\mu\nu\rho\sigma}\left(V_{\mu\nu\rho\sigma}\right)^* = \frac{16\kappa^2}{3}V^2~,
\end{equation}
with 
\begin{equation}
\label{vertex-squared}
    \begin{split}
        V^2 &= 10[(q_1\cdot q_2)^2(q_3\cdot p)^2 + (q_1\cdot q_3)^2(q_2\cdot p)^2 +\\
        &+ (q_2\cdot q_3)^2(q_1\cdot p)^2] - 8[(q_1\cdot q_2)(q_3\cdot p)(q_1\cdot q_3)(q_2\cdot p) +\\
        &+ (q_1\cdot q_2)(q_3\cdot p)(q_2\cdot q_3)(q_1\cdot p) +\\
        &+ (q_1\cdot q_3)(q_2\cdot p)(q_2\cdot q_3)(q_1\cdot p)] +\\
        &+ 6[(q_1\cdot q_2)(q_1\cdot q_3)(q_2\cdot q_3)p^2 + (q_1\cdot q_2)(q_1\cdot p)(q_2\cdot p)q_3^2 +\\
        &+ (q_1\cdot q_3)(q_1\cdot p)(q_3\cdot p)q_2^2 + (q_2\cdot q_3)(q_2\cdot p)(q_3\cdot p)q_1^2] +\\
        &+ 2[(q_1\cdot q_2)^2q_3^2p^2 + (q_3\cdot p)^2q_1^2q_2^2 + (q_1\cdot q_3)^2q_2^2p^2 +\\
        &+ (q_2\cdot p)^2q_1^2q_3^2 + (q_2\cdot q_3)^2q_1^2p^2 + (q_1\cdot p)^2q_2^2q_3^2] + \\
        &+ 3q_1^2q_2^2q_3^2p^2~.
    \end{split}
\end{equation}
In the $Z^0$ rest frame, we have
\begin{align}
    p = (m_Z,\vec{0}) \quad\text{and}\quad q_i = (E_i,\vec{q_i})~,
\end{align}
which, along with momentum conservation, allows us to write all of the 4-momentum relations in (\ref{vertex-squared}) in terms of the final states energies:
\begin{align}
    p\cdot q_1 = m_ZE_1 \;\text{and}\; q_2\cdot q_3 = \frac{m_Z^2-m_X^2}{2} - m_ZE_1~;\\
    p\cdot q_2 = m_ZE_2 \;\text{and}\; q_1\cdot q_3 = \frac{m_Z^2-m_X^2}{2} - m_ZE_2~;\\
    p\cdot q_3 = m_ZE_3 \;\text{and}\; q_1\cdot q_2 = \frac{m_Z^2-m_X^2}{2} - m_ZE_3~.
\end{align}

With those relations established, we can write the differential decay width of $Z^0$ into $3X$:
\begin{equation}
    d\Gamma= \frac{1}{3!}\frac{1}{2 M_Z} \langle|\mathcal{M}|^2\rangle d\Pi_3~,
\end{equation}
with the three-body phase space $d\Pi_3$ defined as
\begin{equation}
\begin{split}
    d\Pi_3 &= \frac{d^3\mathbf{q}_1}{(2\pi)^3 2E_1}\frac{d^3\mathbf{q}_2}{(2\pi)^3 2E_2}\frac{d^3\mathbf{q}_3}{(2\pi)^3 2E_3}\times\\&\times(2\pi)^4\delta^4(p-q_1-q_2-q_3)~.
\end{split}
\end{equation}

Integrating over $\mathbf{q_3}$, we can make use of the delta function, which guarantees energy and 3-momentum conservation, to express $\mathbf{q_3}=-(\mathbf{q_1}+\mathbf{q_2})$. The energy $E_3$ can then be rewritten as
\begin{eqnarray}
\label{E3}
        &&E_3 = \sqrt{|\mathbf{q_1}|^2+|\mathbf{q_2}|^2+2|\mathbf{q_1}||\mathbf{q_2}|\cos\theta_{12}+m_X^2}\\
        &&=\sqrt{E_1^2+E_2^2+2\sqrt{(E_1^2-m_X^2)(E_2^2-m_X^2)}\cos\theta_{12}-m_X^2}~,\nonumber
\end{eqnarray}
with $\cos\theta_{12}$ being the angle between $\mathbf{q}_1$ and $\mathbf{q}_2$. Defining $\mathbf{q}_1$ along the z-axis and using that
\begin{align}
    \frac{d^3\mathbf{k}}{2E}=\frac{\frac{1}{2}|\mathbf{k}|d(\mathbf{k}^2)d(\cos\theta)d\phi}{2E} = \frac{1}{2}|\mathbf{k}|dEd(\cos\theta)d\phi~,
\end{align}
where we used the fact that $d(E^2)=d(\mathbf{k}^2+m^2)=d(\mathbf{k}^2)$. the rest of the phase space can be recast into the form
\begin{equation}
\begin{split}
\label{delta-arg}
        d\Pi_3 &= \frac{|\mathbf{q_1}||\mathbf{q_2}|}{8(2\pi)^5E_3}dE_1d(\cos\theta)d\phi dE_2d(\cos\theta_{12})d\phi'\times\\ &\times\delta(m_Z-E_1-E_2-E_3(\cos\theta_{12}))~,
\end{split}
\end{equation}
\\
Integrating over $\theta$, $\phi$, $\phi'$ and making use of the composition property of the delta function, we are left with
\begin{equation}
\begin{split}
        d\Pi_3 &= \frac{\pi^2}{(2\pi)^5}dE_1dE_2d(\cos\theta_{12})\delta(\cos\theta_{12} - \cos\theta_0)~,
\end{split}
\end{equation}
in which $\cos\theta_0$ is the the root of the delta argument in (\ref{delta-arg}). The kinematics of a three body decay, constrained by the momentum conservation imposed by delta function, limit the range of $E_1$ and $E_2$ to be $\left(\frac{m_Z-m_X}{2}-E_2,\frac{m_Z-m_X}{2}\right)$ and $\left(m_X,\frac{m_Z-m_X}{2}\right)$, respectively. Integrating over the remaining variables, we find the decay width:
\begin{equation}
\label{decay-width}
\begin{split}
    \Gamma = &\frac{\kappa^2}{72\pi^3} \frac{1}{960} m_Z \left(4 m_Z^8 - 6 m_X m_Z^7 + 69 m_X^2 m_Z^6\right.\\ 
    &- 220 m_X^3 m_Z^5 -545 m_X^4 m_Z^4 + 3530 m_X^5 m_Z^3 \\
    &- \left.6693 m_X^6 m_Z^2 +1448 m_X^7 m_Z -5235 m_X^8\right).
\end{split}
\end{equation}

\begin{figure}[h!]
    \centering
    \includegraphics[scale=0.65]{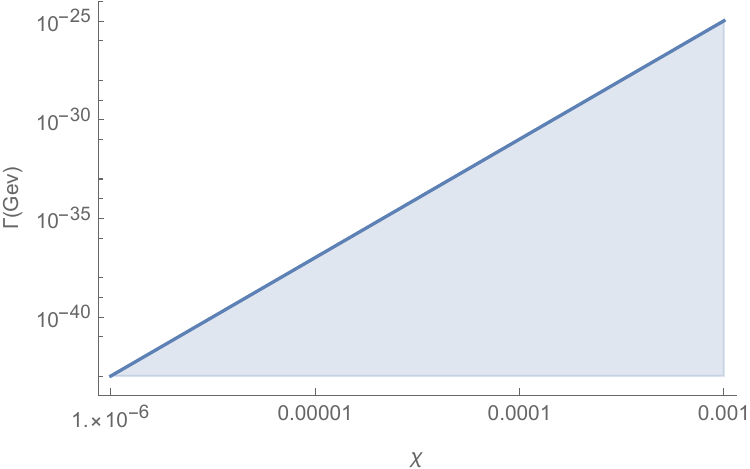}
    \caption{Decay width of $Z^0\to3X$ as a function of the mixing parameter $\chi$.}
    \label{graph}
\end{figure}

We can utilize the estimates on the kinetic mixing parameter $\chi$ and on the the Born-Infeld energy scale, defined by $\beta$, to set some limits on the decay width $\Gamma$. From the X-boson phenomenology \cite{x-boson} and the bound on the Born-Infeld parameter $\beta$ \cite{beta-constraint} found in the literature:
\begin{equation}\label{beta-chi}
10^{-6}<\chi<10^{-3}\;\text{and}\;\sqrt{\beta} \gtrsim 90\;\text{GeV}~,
\end{equation}
along with the masses $m_X=16.7$ MeV and $m_Z=91.2$ GeV for the X and $Z_0$ bosons, respectively, and the already established Weinberg angle $\theta_W$, we can estimate $\Gamma$ to be bounded from above by the curve:
\begin{equation}
\label{gamma-chi}
    \Gamma(\chi) = 3.65\cdot10^{-7} \chi^6 \left(\chi^2+0.53\right)^2\; \text{GeV}~,
\end{equation}
where we considered the BI-parameter $\beta$ value as its lower bound, {\it i.e.} $\sqrt{\beta}=90$ GeV. Since $\Gamma \propto 1/\beta^4$, it decreases as $\beta$ increases. Therefore, the allowed values will always lie below the curve given by (\ref{gamma-chi}), represented in Figure \ref{graph} by the shaded region in light blue.

The allowed values of the kinetic mixing $\chi$ established places the presented model within the range of WIMP (Weakly Interacting Massive Particle) and FIMP (Feebly interacting Massive Particle) dark matter, which accepts scenarios with kinetic mixing parameters well below the ones utilized here, as shown in \cite{dark-photon-constraint}, being still capable of satisfying some of the constraints imposed over the model by experimental and observational data. In particular, it has a wide range of agreement with the latest NA64 collaboration results \cite{na64}, which best accommodates the mass of the X17 within its mass range for the dark photon. 

\textit{The $Z^0\to 3\gamma$ decay}---. However, the channel which shows the most promising results is the $Z^0$ decay into three photons. Apart from the constant $\kappa$, the vertex factor is given by the same expression (\ref{vertex_factor}), now with constant $\zeta$:
\begin{equation}
    \zeta =  \frac{\sin{2\theta_W}\cos^2{\theta_W}}{16\beta^2}\left(\chi^3 - 1\right),
\end{equation}

The following steps are similar, the difference being that the final states are now all massless, which alters the expression for $V^2$ in the unpolarized squared amplitude (since $q_1^2=q_2^2=q_3^2=0$) and the kinematical constraints of the three body decay, limiting the energy ranges for $E_1$ and $E_2$ in the final integration to be $\left(0,\frac{m_Z}{2}\right)$ and $\left(\frac{m_Z}{2}-E_1,\frac{m_Z}{2}\right)$, respectively. This yields the decay width:
\begin{equation}\label{decay-z3g}
    \Gamma = \frac{\zeta^2}{72\pi^3}\frac{m_Z^9}{240},
\end{equation}
which, substituting the $Z^0$ mass, $m_Z$, and the Born-Infeld lower bound, $\sqrt{\beta}$ (\ref{beta-chi}), gives us
\begin{equation}\label{z3g-values}
    \Gamma(\chi) = 3.12\cdot 10^{-7} \left(\chi^3-1\right)^2 \text{ GeV},
\end{equation}
Here we can see a deviation from the strong suppression presented by the previous decay. This more significant contribution comes from the low values of the mixing parameter $\chi$ due to the weak coupling between $U_Y(1)$ and $U(1)_{B-L}$ being less influential in the decay, since it does not contribute to the photon field, which is why the leading contribution does not depend on $\chi$.

However, this rare decay has a clear experimental bound for its branching ratio, which implies that we could use expression (\ref{decay-z3g}) to estimate the value the Born-Infeld parameter $\sqrt{\beta}$, instead of using the value derived from light-by-light scattering.
From the experimentally settled values for the total decay width of the $Z^0$ \cite{z-width} and the branching ratio upper bound for the $Z \to \gamma\gamma\gamma$ decay \cite{z-br}:
\begin{equation}
    \Gamma_Z = 2.49~ \text{GeV}\; \text{ and }\; \text{BR}(Z \to \gamma\gamma\gamma) < 2.2\times 10^{-6},
\end{equation}
together with the decay width (\ref{decay-z3g}), we get a upper bound for the BI-parameter of:
\begin{equation}
    \sqrt{\beta}  \gtrsim 62.9 ~\text{GeV},
\end{equation}
which is a lower value in comparison with the pure hypercharge calculations \cite{bi-hypercharge2} and the light-by-light scattering \cite{beta-constraint}. Here we ignored the corrections coming from the mixing parameter $\chi$ to the decay width, since its contribution is negligible even for the highest value. However, the exclusion of parity odd term, such as $Y_{\mu\nu}\tilde{Y}^{\mu\nu}$, from the Born-Infeld lagrangian could be the reason for this lower value, since its contribution is neglected by the present analysis.

We can also see that the branching ratio for this event, neglecting the $\chi$ corrections:
\begin{equation}
    \text{BR}(Z \to \gamma\gamma\gamma) = 1.25 \times 10^{-7},
\end{equation}
using the $\gamma\gamma\to\gamma\gamma$ scattering value for $\sqrt{\beta}$, that is, dividing equation (\ref{z3g-values}) by the total decay of the $Z^0$, comes much closer to the experimental bound of $2.2\times 10^{-6}$ than the results from SM calculations, where the contributions begin at one loop, with a rate in the order of $10^{-10}$ (cf. \cite{bi-hypercharge2}). At the same time, it still respects the experimental upper bound, maintaining itself as a viable candidate for the description of this decay.

\textit{Conclusions}---. In conclusion, it should also be emphasized that the feeble contribution of the $Z^0\to 3X$ decay to the total decay width of the $Z^0$, which we already expected to be small due to the small mixing $\chi$ is further suppressed by the Born-Infeld $\beta$ parameter. The advantage of considering a non-linear extension is the presence of the $Z^0 \to 3X$ decay already at tree-level, with no need to consider loop order contributions, due to the wider range of interactions allowed by (\ref{bi-nl}). Also, it is a way of investigate how a gauge extension of the standard model may affect the current bound for non-linearity in the standard model \cite{beta-constraint}. For this purpose, we also investigated the SM allowed decay of $Z\to3\gamma$, which in presented better results in an gauge extended model than the previous SM results obtained via a Born-Infeld hypercharge sector, comparing both to the light-by-light scattering results. The next steps include the investigation of $Z^0$ decays with mixed $X,\gamma$ final states, like the $Z^0\to X\gamma\gamma$ decay. These could be subject of investigation in the future Z factories \cite{dm-z-factory} and potentially probe into new hidden-sector particles.

Although its detection as by-products of colliders is unlikely, this framework could be useful to explain some highly energetic astronomical events, such as gamma ray bursts (since decays involving photons also appear as quartic vertices from the expansion of the lagrangian (\ref{bi-nl})), many of which remain without an explanation of where it were originated. The proposal of the X17 as a light mass dark photon candidate can be inserted in the search for MeV energy range gamma-ray searches \cite{mev-gamma-ray,mev-gamma-ray2}, still largely unexplored, albeit being the peak emissivity range for a number of astrophysical events, as the already mentioned gamma-ray bursts and objects such as blazars and pulsars, which motivates the investigation of the X17 decay into photons, also present in (\ref{bi-nl}), which will be subject of a future paper together we the ones already mentioned.

\textit{Acknowledgements}---. The authors dedicate this work to the 60th birthday of Prof. Daniel H.T. Franco. The authors would like to thank Professor P. Fayet for the many discussions and suggestions, they are also indebted to the anonymous referee for valuable comments and suggestions on the text. CAPES-Brazil is acknowledged for invaluable financial help.


\begin{references}

\bibitem[]{fayet1} P. Fayet, Nucl. Phys. B 187 (1981) 184.

\bibitem[]{fayet7} P. Fayet, Phys. Lett. B 172 (1986) 363.

\bibitem[]{fayet4}  P. Fayet, Phys. Lett. B 227 (1989) 127.

\bibitem[]{fayet5} P. Fayet, Nucl. Phys. B 347 (1990) 743.

\bibitem[]{fayet6} C. Boehm, P. Fayet, Nucl.Phys.B 683 (2004) 219-263

\bibitem[]{fayet2} C. Bouchiat, P. Fayet, Phys. Lett. B 608 (2005) 87.

\bibitem[]{fayet3} P. Fayet, Eur. Phys. J. C 77 (2017) 53.

\bibitem[]{atomki1} A.J. Krasznahorkay \textit{et al}., Phys. Rev. Lett. 116 (2016) 042501.

\bibitem[]{x-model1} J.L. Feng \textit{et al}., Phys. Rev. Lett. 117 (2016) 071803.

\bibitem[]{x-model2} P.H. Gu, X.G. He, Nucl. Phys. B 919 (2017) 209.

\bibitem[]{8be-alt} X. Zhang, G.A. Miller, Phys. Lett. B 773 (2017) 159.

\bibitem[]{4he-alt} M. Viviani \textit{et al}., Phys. Rev. C 105 (2022) 014001.

\bibitem[]{atomki2} A.J. Krasznahorkay \textit{et al}., Phys. Rev. C 104 (2021) 044003.

\bibitem[]{atomki3} A.J. Krasznahorkay \textit{et al}., Phys. Rev. C 106 (2022)  L061601.

\bibitem[]{besiii} J. Jiang, L.B. Chen, Y. Liang, C.F. Qiao, Eur. Phys. J. C 78 (2018) 456.

\bibitem[]{x-boson-search2} T. Bandyopadhyay, S. Chakraborty, S. Trifinopoulos, J. High Energ. Phys. 2022 (2022) 141.

\bibitem[]{belle-bes} K. Ban \textit{et al}., J. High Energ. Phys. 2021 (2021) 91.

\bibitem[]{x-boson-search1} D. Banerjee \textit{et al}. (NA64 Collaboration), Phys. Rev. Lett. 120 (2018) 231802.

\bibitem[]{babar} J.P. Lees \textit{et al}. (BaBar Collaboration), Phys. Rev. Lett. 119 (2017) 131804.

\bibitem[]{megii} K. Afanaciev \textit{et al.} (MEG-II Collaboration), arXiv:2411.07994 [nucl-ex].

\bibitem[]{padme1} E. Nardi, C.D.R. Carvajal, A. Ghoshal, D. Meloni, M. Raggi, Phys. Rev. D 97 (2018) 095004.

\bibitem[]{padme2} L. Darmé, M. Mancini, E. Nardi, M. Raggi, Phys. Rev. D 106 (2022) 115036.

\bibitem[]{padme3} S. Bertelli \textit{et al.} (PADME Collaboration), J. High. Energ. Phys. 06 (2025) 040.  

\bibitem[]{newjedi} B. Bastin \textit{et al.} (New JEDI Collaboration), EPJ Web Conf. 275 (2023) 01012.

\bibitem[]{icecube} R. Enberg, Y. Hiçyılmaz, S. Moretti, C. P. de los Heros, H. Waltari, J. High. Energ. Phys. 06 (2025) 182.

\bibitem[]{compobj} A. Kanakis-Pegios, V. Petousis, M. Veselský, Jozef Leja, Ch. C. Moustakidis, Phys. Rev. D 109 (2024) 043028.

\bibitem[]{anomaly} A. Capolupo, A. Quaranta, R. Serao, Physics of the Dark Universe 47 (2025) 101748.

\bibitem[]{cern} C. Gustavino \textit{et al.}, Nucl. Instrum. Meth. A 1072 (2025) 170087.

\bibitem[]{decays} L.T.T. Uyen, F.F. Lee., G.L. Lin, PoS ICHEP2024 (2025) 261.

\bibitem[]{libe} P. Gysbers, P. Navrátil, K. Kravvaris, G. Hupin, S. Quaglioni, Phys. Rev. C 110 (2024) 1, 015503.

\bibitem[]{hhehh} M. Viviani, E. Filandri, L. Girlanda, C. Gustavino, A. Kievsky, L.E. Marcucci, Phys. Rev. C 111 (2025) 3, 034002.

\bibitem[]{review1} D. Barducci, C. Toni, J. High Energ. Phys. 2023, 154 (2023), arXiv:2212.06453v1.

\bibitem[]{review2} D.S.M. Alves, D. Barducci, G. Cavoto  \textit{et al}., Eur. Phys. J. C 83, 230 (2023).

\bibitem[]{review3} A. J. Krasznahorkay \textit{et al.}, Universe 10 (2024) 11, 409

\bibitem[]{review4} C. Gustavino, Universe 10 (2024) 7, 285

\bibitem[]{x-boson} M.J. Neves, E.M.C. Abreu, J.A. Helay\"el-Neto, Acta Phys. Pol. B 51 (2020) 909.

\bibitem[]{stueckelberg-mech} C. Han, M.N. López-Ib\'a\~nez, B. Peng, J.M. Yang, Nucl. Phys. B 959 (2020) 11515.

\bibitem[]{dm-z-factory} J. Liu \textit{et al}., Int. J. Mod. Phys. A 34 (2019) 1940010.

\bibitem[]{atlas-gamma-gamma} The ATLAS collaboration., Aad, G., Abbott, B. et al. J. High Energ. Phys. 2021, 243 (2021)

\bibitem[]{bi-hypercharge1} M.J. Neves, L.P.R. Ospedal, J.A. Helay\"el-Neto, P. Gaete, Eur. Phys. J. C 82 (2022) 327.

\bibitem[]{bi-hypercharge2} P. De Fabritiis, P.C. Malta, J.A. Helay\"el-Neto, Phys. Rev. D 105 (2022) 016007.

\bibitem[]{modmax1} P. Gaete, J. Helayël-Neto, Eur. Phys. J. C 74 (2014) 2816.

\bibitem[]{modmax2} S.I. Kruglov, Eur. Phys. J. C 75 (2015) 88.

\bibitem[]{log1} I. Bandos, K. Lechner, D. Sorokin, and P. T. Townsend, Phys. Rev. D 102 (2020) 121703(R).

\bibitem[]{log2} D. Sorokin, Fortschr. Phys. 70 (2022) 2200092.

\bibitem[]{log3} K. Lechner, P. Marchetti, A. Sainaghi, and D. Sorokin, Phys. Rev. D 106 (2022) 016009.

\bibitem[]{beta-constraint} J. Ellis, N.E. Mavromatos, T. You, Phys. Rev. Lett. 118 (2017) 261802.

\bibitem[]{dark-photon-constraint} M. Goodsell \textit{et al}., J. High Energ. Phys. 11 (2009) 027.

\bibitem[]{na64} C. Cazzaniga \textit{et al}. (NA64 Collaboration), Eur. Phys. J. C 81 (2021) 959.

\bibitem[]{z-width} P.A. Zyla \textit{et al}. (Particle Data Group), Prog. Theor. Exp. Phys. 2020, 083C01 (2020) and 2021 update.

\bibitem[]{z-br} ATLAS Collaboration, Eur. Phys. J. C 76 (2016) 210 .

\bibitem[]{mev-gamma-ray} A. De Angelis \textit{et al}., Exp. Astron. 51 (2021) 1225. 

\bibitem[]{mev-gamma-ray2} J. Guo, L. Wu, B. Zhu, Phys. Lett. B 840 (2023) 137853



\end{references}
\end{document}